\documentclass[prb,onecolumn,showpacs,10pt,nofootinbib]{revtex4}
\usepackage{graphicx}
\usepackage{amsmath}
\usepackage{amssymb}
\begin{document}
\title{TRANSPORT PROPERTIES IN THE ``STRANGE METAL PHASE'' OF HIGH $T_c$
CUPRATES: SPIN-CHARGE GAUGE THEORY VERSUS EXPERIMENTS}

\author{P. A. Marchetti}
\affiliation{Dipartimento di Fisica ``G. Galilei'', INFN, I-35131
Padova, Italy}
\author{ G. Orso}
\affiliation{International School for Advanced Studies (SISSA), INFM,
Via Beirut,  34014 Trieste, Italy}
 \author{Z. B. Su}
 \author{  L. Yu}
 \affiliation{Institute of Theoretical Physics and Interdisciplinary Center of Theoretical Studies,\\
 Chinese Academy of Sciences, 100080 Beijing, China;\\
 Center for Advanced Study, Tsinghua Unievrsity, 100084 Beijing, China}

\date{\today}
 \begin{abstract}
The SU(2)$\times$U(1) Chern-Simons spin-charge gauge approach
developed earlier to describe the transport properties of the
cuprate superconductors in the ``pseudogap'' regime, in
particular, the metal-insulator crossover of the in-plane
resistivity, is generalized to the ``strange metal'' phase at
higher temperature/doping. The short-range antiferromagnetic order
and the gauge field fluctuations, which were the  key ingredients
in the  theory for the pseudogap phase, also play an important
role in the present case.  The main difference between these two
phases is caused by the existence of an underlying statistical
$\pi$-flux lattice for charge carriers in the former case, whereas
the background flux is absent in the latter case. The Fermi
surface then changes from small ``arcs''  in the pseudogap to a
rather large closed line in the strange metal phase. As a
consequence the celebrated linear in $T$ dependence of the
in-plane and out-of-plane resistivity is explicitly derived. The
doping concentration and temperature dependence of theoretically
calculated in-plane and out-of-plane resistivity, spin-relaxation
rate and AC conductivity are compared with experimental data,
showing good agreement.

 \end{abstract}

\maketitle
\section{Introduction: the ``Strange Metal Phase''}
In spite of the intensive studies on High T$_c$ superconductors
for almost two decades, the phase diagram of this fascinating
system is still under debate.\cite{norman} According to the RVB
scenario,\cite{bas} the temperature-doping concentration plane is
divided into four regions: The pseudogap (PG), the strange metal
(SM), the superconducting (SC) state and the  Fermi liquid  state.
An alternative is the quantum critical point (QCP)
scenario\cite{sachdev} assuming the existence of a QCP under the
SC dome controlling the behavior of the system in a rather large
quantum critical regime. In a sense that regime corresponds to the
SM phase in the RVB picture, the difference being that the PG
phase in the QCP scenario has a true long range order, and the
boundary between PG and SM phases is a phase transition line,
whereas in the RVB picture it is a crossover and it coincides with
the superconducting transition line at high dopings. There are
several theoretical as well as ``experimental'' proposals on the
origin of the
QCP.\cite{sachdev-rmp,kivelson,varma,castellani,pines,abanov,loram}

Recently, we have developed a spin-charge gauge approach to
describe the PG phase in cuprate superconductors, particularly
focusing on the metal-insulator crossover (MIC)
phenomena.\cite{Mar,Dai,magneto,mdosy,anis}  Within a unified
framework we have calculated the in-plane and out-of-plane
resistivity, including the effect of external magnetic field,
optical conductivity, nuclear magnetic resonance (NMR) relaxation
rate and the spectral weight of the electron Green's function,
finding a good agreement with experimental data. In this approach
based on spin-charge decomposition applied to the 2D $t$-$J$
model, the spinon dynamics is described by a nonlinear
$\sigma$-model with a theoretically derived mass gap $m_s \sim J
(\delta |\ln \delta|)^{1/2}$, where $J$ is the exchange integral,
$\delta$ the doping concentration. The holon is fermionic with
``small'' Fermi surface  $(\epsilon_F \sim t \delta)$ (with $t$ as
the hopping integral) centered around $(\pm \pi/2, \pm \pi/2)$ in
the Brillouin zone and a ``Fermi-arc'' behavior for the spectral
weight. Both holons and spinons are strongly scattered by gauge
fluctuations. As an effect of these gauge interactions, the spinon
mass picks up a dissipative term: $m_s \rightarrow M_{T}=(m_s^2- i
c {T/ \chi})^{1/2}$, where $\chi \sim t \delta^{-1}$ is the
diamagnetic susceptibility and $c$ a numerical constant. This
shift in turn introduces a dissipation in the spinon-gauge sector,
whose behavior dominates the low energy physics of the system. The
competition between the mass gap and the dissipation is
responsible for the MIC, giving rise to a broad peak in the DC
conductivity. At low temperatures the antiferromagnetic (AF)
correlation length $\xi \sim m_s^{-1}$ is the determining scale of
the problem, leading to a insulating behavior, while at higher
temperatures, the de Broglie wave length $\lambda_T \sim
(\chi/T)^{1/2}$ becomes comparable, or even shorter than $\xi$,
giving rise to metallic conductivity.

 Since this theory was originally \cite{Mar,mdosy}  formulated to describe the
MIC  in the in-plane resistivity $\rho_{ab}$ and related phenomena
taking place in the PG ``phase'',  its range of applicability was
therefore limited to
 underdoped systems and low temperatures where the AF correlation
 length $\xi$ is smaller or of the order of the thermal de Broglie
wavelength $\lambda_T $.  In fact, our theory correctly describes
the low-temperature insulating behavior and MIC up to the
inflection point, where the second derivative of $\rho$ {\it
w.r.t.} temperature vanishes. However, the ``high temperature
asymptotics'' derived from the theory $\rho \sim T^{1/4}$ is not
correct.  At higher temperatures $T \sim T^*$, the PG temperature,
 underdoped cuprates crossover to a new, SM ``phase''.
Overdoped cuprates also reach this phase increasing $T$, but
presumably from an ordinary Fermi liquid state. The SM phase is
{\it metallic} in nature with anomalous yet rather simple
temperature dependence of physical observables such as the
celebrated $T-$linearity of both in-plane and out-of-plane
resistivity at sufficiently high temperatures.

Experimentally, the SM phase shares with the PG phase an
 AF short range order   which in our
approach originates from the gapful spin excitations due to
scattering against spin vortices attached to the moving holes.
There are strong indications, in particular from the
angle-resolved photoemission spectroscopy (ARPES)
experiments,\cite{largeFS} that charge degrees of freedom undergo
 a radical rearrangement near the crossover temperature
$T^*$: the excitations far from the zone diagonals, {\it i.e.},
located near $(\pi,0)$ of the Brillouin zone, become gapless and
lead to a {\it large} closed Fermi surface. This means that the
density of {\it effective} charge carriers has grown from
$\delta$, characteristic of the PG phase, to $1-\delta$, the value
expected from the band structure calculations.

The transport and optical properties in the SM phase were
considered earlier by various theoretical treatments, including
the slave-particle mean field theory,\cite{fukuyama} the gauge
field approach,\cite{Lee,iowi,ichinose} ``nearly AF Fermi liquid''
theory,\cite{stojkovic} and the closely related spin-fermion
model.\cite{abanov} In the last approach a special role is
assigned to the ``hot spots'', {\it i.e.}, those points on the
Fermi surface, separated by the AF vector ($\pi,\pi$) from their
counterparts. As a kind of opposite limit, the ``cold spots''
(intersections of Fermi surface with zone diagonals) model was
also considered,\cite{io-mi} where these ``cold spots'' are argued
to dominate the in-plane and out-of-plane transport properties.
However, it is fair to say that a complete understanding of the
diversified experimental findings in this unusual phase is still
lacking.

In this paper we generalize our spin-charge gauge approach to
consider the SM phase, and compare the calculated transport
properties with experimental data in this regime. The AF short
range order and the gauge field fluctuations which were the key
ingredients in the theory for the PG phase still play an important
role here. The main difference, however, is the following: In the
PG phase near half filling there is an underlying statistical
$\pi$-flux lattice for holons. As the doping (or temperature)
increases, this $\pi$-flux lattice is more and more disturbed, and
at certain point it ``melts''. The energetically favorable
configuration should then correspond to zero flux, instead of
$\pi$-flux ({\it Assumption} 0). It turns out that our spin-charge
gauge approach can be generalized to this case to describe the
large Fermi surface instead of ``Fermi-arcs''.
 As a consequence the celebrated linear in $T$ dependence of in-plane
and out-of-plane resistivity is  explicitly derived. Since the
formalism is similar, we will only sketch the major steps,
referring the reader to our previous papers Refs.
\onlinecite{Mar,mdosy} for more detail. The rest of the paper is
organized as follows: the spin-charge gauge approach is outlined
in Sec. II, while the effects of the gauge field are considered in
Sec. III.  The comparison of theoretical results with experimental
data on in-plane and out-of plane resistivity,  NMR relaxation
rate and AC conductivity is presented in Sec. IV with some brief
concluding remarks in Sec. V.  The proof that one can satisfy the
{\it Assumption} 0 for the free energy minimum is given in the
Appendix.

     Readers only interested in the final theoretical results for the SM
     phase and comparison with experiment can skip Secs. II and
     III, and go directly to Sec. IV.

\section{Spin-charge gauge approach}

As mentioned above,  to discuss the ``SM phase'' we follow the
same strategy as that  adopted for the ``PG phase''. Therefore we
outline  the changes instead of repeating the whole procedure. The
discussion however will be sufficiently complete that using Refs.
\onlinecite{Mar,mdosy} as a guide the interested reader will have
no difficulty to fill in the missing steps. From now on we adopt
the lattice constant $a=1$.

We use the 2D $t$-$J$ model  to describe the CuO layers in high
$T_c$ cuprates  and we treat it in an ``improved Mean Field
Approximation''(MFA)  via a gauge theory of spin-charge
decomposition, obtained by
 gauging the global spin and charge symmetries of the model by
  introducing Chern-Simons (C-S) gauge fields.
To the fermion $c$ of the gauged model we apply the spin-charge
decomposition: $c \sim  H z_\alpha$, where $H$ is the holon, a
charged spinless fermion, and $z_\alpha$ is the spinon, neutral
spin $1/2$ boson of a nonlinear $\sigma$ model. To derive the MFA
a key step is to find a ``reference spinon configuration'' {\it
w.r.t.} which one expands the fluctuations described by
$z_\alpha$. Such reference configuration was found by optimizing
the free energy of holons in a  self-consistently fixed
holon-dependent spinon background, in a kind of Born-Oppenheimer
approximation. In Ref. \onlinecite{mdosy} it has been shown that a
lower bound on this free energy  can be obtained via the
optimization of the free energy of a gas of spinless holes with
the same density of holons, on a square lattice, with n.n. hopping
parameter given by $t U_{<ij>}$. Here $U_{<ij>}$ is a
time-independent complex gauge field related to the C-S charge and
spin gauge fields, denoted by $B$ and $V$, respectively, by
\begin{equation}
 U_{<ij>} \sim  e^{ -i \int_{<ij>}B }
(\sigma_x^{|i|} (P e^{ i\int_{<ij>} V} )\sigma_x^{|j|})_{11},
\label{iden}
\end{equation}
where $P$ denotes the path-ordering, $\sigma_x$ the Pauli matrix
and $|i|=0$ if $i$ is a site on the N\'eel
 sublattice containing the origin and $|i|=1$ if $i$ is on the other N\'eel sublattice.

At low temperature and small doping concentration, {\it i.e.} in
the parameter region to be compared with the ``PG phase'' of the
cuprates, it has been argued in Ref. \onlinecite{Mar} that the
optimal configuration of $U$ carries a flux $\pi$ per plaquette
and it has been shown that one can find configurations of $B$ and
$V$ reproducing this behavior on average. Then neglecting the
feedback of holon fluctuations on $B$ and that
  of spinon fluctuations on $V$,
the charge C-S gauge field $B$ carries flux $\pi$ per plaquette
and via Hofstadter mechanism converts the spinless holon $H$ into
a Dirac fermion with a small Fermi surface centered at the four
nodes $\Bigl(\pm \frac{\pi}{2}, \pm \frac{\pi}{2}\Bigr)$; the spin
C-S gauge field $V$ dresses the holons with spin vortices and the
spinons in this gas of ``slowly moving dressed holons'' acquire a
mass $m_s \sim J\sqrt{|\delta \ln \delta|}$. The concentration
dependence of the AF correlation length $\xi \sim m_s^{-1}$
derived from the theory for low dopings is consistent with what
was determined from the neutron experiments.\cite{Birgenau}

Dirac holons and massive spinons are coupled by a $U(1)$
slave-particle gauge field $A$, called $h/s$ in Ref.
\onlinecite{Mar}. The low energy effective action for $A$ is
obtained upon integration of the matter fields and since
 holons have a non-vanishing density at the Fermi surface, it exhibits a Reizer singularity
\cite{Reizer}   dominating  at large scales: for small $q,\omega,
\omega/|\vec q|$

\begin{equation} \label{pi}
\langle A^T A^T \rangle (\omega,\vec q) \sim ( \chi |\vec q |^2 +
i \kappa {\omega / |\vec q|})^{-1},
\end{equation}
where $\chi \sim \delta^{-1}$ is the diamagnetic susceptibility
and $\kappa \sim \delta$ the Landau damping. In turn the Reizer
singularity produces  a shift in the mass of spinons $ m_s
\rightarrow M=(m_s^2- i c {T / \chi})^{1/2}$, as the leading
effect \cite{Dai}, where $c \sim O(1)$ is a positive constant,
thus producing a dissipation, linear in $T$ at low temperatures.
The competition between the two energy scales $m^2_s$ and
$T/\chi$, related to the spinon gap and dissipation is the root in
the spin-charge gauge approach of many crossover phenomena
peculiar to the transport properties of the ``PG phase''.

In our approach, the increase in the density of state at the Fermi
energy $\varepsilon_F$ as we move from the PG to the SM ``phase''
reflects a change in the holon dispersion relation. The $\pi-$flux
statistical field that minimizes the ground state energy near half
filling  was responsible for the small Fermi surface in the PG
phase. Since the SM is metallic in nature, we expect that in this
phase there are no statistical magnetic fields to frustrate the
charge motion and therefore we make the following

\noindent {\it Assumption} 0: In the optimal configuration (for
the range of parameters corresponding to the SM phase) the flux
per plaquette carried by $U$ is 0, i.e.
\begin{equation}\label{0-flux}
\arg(U_{\partial p})=0.
\end{equation}

In favor of this assumption we can offer the following argument:
It has been rigorously proved by Lieb\cite{lieb} that at half
filling the optimal configuration for a magnetic field on a square
 lattice is translationally invariant, with flux $\pi$ for each plaquette at arbitrary high temperature.
  On the other hand, it is well known that at low density and high temperature the optimal configuration
   has zero flux   per plaquette. At zero temperature it has been proved that the ground state energy has
    a minimum corresponding to one flux quantum per spinless fermion.\cite{rammal} Numerical simulations
    suggest that increasing the temperature gives rise to a competition between these minima and at high
     temperatures only the zero flux  and $\pi$-flux  configurations survive.\cite{qin} Therefore
     one can argue that at
     sufficiently high doping and temperature {\it Assumption} 0 is satisfied. By analyticity it is then
      reasonable to assume it holds in the entire SM phase except near the crossover
      boundaries to other ``phases''.
Taking {\it Assumption} 0 for granted, we show explicitly in the
Appendix that (under the same approximation as used for the PG
phase) one can find a gauge field configuration that satisfies it
on average,
 so that our spinon field $z_\alpha$ will describe the fluctuations around it. The outcome of
 this optimization is that in the SM phase the flux carried by the charge gauge field
 responsible for the statistical flux in the PG phase is cancelled by the
  flux carried by the spin gauge field on average.

The flux change does not affect the spinon action, therefore it is
formally identical to that considered for the PG case, thus
leading to the same low energy effective action\cite{tree}
\begin{equation}
   S = \int d^{3}x  \frac{1}{g}
   \left[ v_{s}^{-2} \left| \left( \partial_{0} - i A_{0} \right) z_{\alpha}
   \right|^{2} + \left| \left( \partial_{\mu} - i A_{\mu}
   \right) z_{\alpha} \right|^{2} + m_{s}^{2} z_{\alpha}^{*}
   z_{\alpha}
   \right],
\end{equation}
 where $g \sim J^{-1},\, v_s \sim J\,, x=(v_sx^0,\vec{x}),\, A =(v_sA_0,\vec{A})$.
 This implies that in our approach the AF short range order is  characteristic of both  PG
 and  SM phases; however we will see that in the SM phase it is less effective.
The action derived above is for $T=0$ and  it can be argued to be
correct only for temperature smaller
 than the mass gap; one can improve the situation by approximately taking into account finite
  temperature
  correction by including in the action the spinon thermal mass term found in the renormalized classical
region of the model,\cite{belk} with thermal mass  $m_T \sim (2\pi
\rho_s/J) e^{-2 \pi \rho_s /T}$,
  where $\rho_s$
  is the renormalized spin stiffness. If we adopt for $2 \pi \rho_s$ the value (150 meV) used
   in  fitting
   the inverse magnetic correlation length of underdoped samples of LSCO in Ref. \onlinecite{belk},
  the order of magnitude and the qualitative temperature and doping dependence of the experimental
  data are reproduced  using the above derived  formula $ m_s(T)=(m_s^2+m_T^2)^{1/2}$.
   The renormalized classical formula holds only for $T << 2 \pi \rho_s$ and it has been
    argued to be correct
    up to  roughly  500 K and this yields an upper limit of validity of the above treatment.

We now turn to holons. The change of statistical flux does affect
deeply the holon motion. Since we would like  to find eventually
the continuum low-energy action, we expand the  holon action  in
powers
 of the lattice constant keeping the leading order, and  the result rewritten in the
 lattice form is:\cite{orso}
\begin{equation}
S_h(H,H^*,A)=\int d^3x\left[ \sum_jH^*_j(i\partial_0-\mu-A_0)H_j+
\sum_{<ij>} (tH_i^*H_je^{i\int_{<ij>}A}+h.c.)\right],
\end{equation}
where $\mu$ is the chemical potential, $A_\mu$ denotes the
slave-particle self-generated $h/s$ gauge field as in the PG
phase. Neglecting gauge fluctuations, the dispersion relation for
holons has changed from the $\pi-$flux phase spectrum
$\epsilon^{PG}(\vec p)=\pm 2t \sqrt{{\cos^2(p_x)}+{\cos^2(p_y)}}$
 restricted to
the magnetic Brillouin zone to the more conventional tight binding
spectrum $\epsilon^{SM}(\vec p)=2t(\cos(p_x)+\cos(p_y))$
 defined in the entire Brillouin zone.
Obviuosly, the bottom of the holon band  is located at the corners
of the Brillouin zone. To write a continuum limit for the holon
action, as a crude approximation we first substitute the Fermi
surface with a circle having the same volume and then make a {\it
particle/hole} conjugation $H_i \rightarrow E_i^*$. Due to this
approximation (and neglecting from beginning the n.n.n. hopping)
all features depending on the detailed structure of the Fermi
surface are clearly lost. The tight binding action defined on the
square lattice is then replaced by a continuum action describing
free particles ($\epsilon_k=k^2/2m^*$) with an effective chemical
potential $\epsilon_F \sim 4t(1-\delta)$, where $-4t$ corresponds
to the bottom of the tight binding band, assuming that the field
$E$ has a well-defined continuum limit. By ``abuse'' of language,
we still call $E$ holon field. The continuum low energy action for
$E$ is given by:
\begin{equation}\label{sstra}
S(E)=\int  d^3x \left[ E^*(x^0,\vec x)(i\partial_0 -\epsilon_F-
A_0-\frac{1}{2m^*} (\vec \nabla  -i\vec A)^2) E(x^0,\vec
x)\right].
\end{equation}

Since  spinons are massive and the charge carriers (holons in the
above sense) described by $E$ are gapless with a finite Fermi
surface, the low-energy effective action for $A$ exhibit a Reizer
behaviour as in the PG phase. Using the above simplified action,
the Landau damping and diamagnetic susceptibility can be easily
evaluated and have a weak dependence on the doping concentration
in the relevant range $\delta \lesssim 0.3 $. First note that $v_F
\sim 2t$ is doping independent, $k_F \propto (1-\delta)$ and
therefore $m^*=k_F/v_F=(1-\delta)/2t$. The parameters entering the
Reizer propagator (\ref{pi}) are then given by
\begin{equation}
\kappa  \sim O(1-\delta),
\end{equation}
\begin{equation}
\chi \sim 1/12 \pi m^* \sim \frac{t}{6 \pi (1-\delta)},
\end{equation}
assuming that the holons give the dominating contribution
(certainly true for sufficiently high $\delta$) to $\chi$.

The main difference {\it w.r.t. } the PG estimates \cite{mdosy} is
that $\delta$ is now replaced by  $(1-\delta)$ so that, roughly
speaking, both quantities vary by a factor of the order $5-10$.

The decrease of the diamagnetic susceptibility implies that the
thermal de Broglie wavelength for holons is shorter {\it w.r.t.}
the PG analogue and therefore the spin-gap effects
($\xi^2<\lambda_T^2$) are less effective, being confined to very
low temperatures. As a result, gauge fluctuations in the spinon
sector strengthen and dominate over the spin gap at all
temperatures down to $T^*$ for the underdoped samples, thus
determining the thermal behavior of many physical properties. In
particular we shall recover some of the distinctive features of
the SM, namely the $T-$linearity of in-plane, out-of-plane
resistivities and spin lattice relaxation time $(T_1T)^{63}$ up to
several hundreds Kelvin.

The low energy effective action obtained in our approach bears
some resemblance with the one derived in the slave boson approach,
\cite{Lee} with however two basic differences: the statistics of
holon and spinon is interchanged and the bosons in our spin-charge
gauge approach are massive, due to the coupling to spin
 vortices absent in the slave boson approach, and ``relativistic''
 due to AF short range order; in the slave boson approach,
 instead, they are non-relativistic and gapless.
In a sense our effective action is closer to the slave fermion
ansatz outlined in Ref. \onlinecite{iowi},
 although the starting point is quite different.

 \section{Gauge field Effects}

In this Section we summarize the physical consequences of the
gauge field fluctuations, expressed in terms of spin and electron
Green's functions.

  The $h/s$ gauge field $A$ renormalizes the
massive spinons in a nontrivial way. The expression obtained for
the {\it dressed} magnon (or spin-wave) correlator   by summing up
the gauge fluctuations via an eikonal approximation obtained for
the PG phase applies here as well:\begin{equation} \label{om1}
   \langle \vec \Omega(x) \cdot \vec \Omega(0) \rangle \sim
   \frac{1}{(x^0)^2-|{\vec x}|^2}
   e^{-2i\sqrt{m_s^2-\frac{ T}{\chi}f(
   \frac{|{\vec x}|Q_0}{2}) }\sqrt{(x^0)^2-{\vec x}^2}
   -\frac{ T}{2\chi}
   Q_0^2 g(\frac{|{\vec x}|Q_0}{2} )
   \frac{(x^0)^2-|{\vec x}|^2 }{m_s^2-\frac{
   T}{\chi}f(\frac{|{\vec x}|Q_0}{2})}} ,
\end{equation}
provided  the new estimates for $\kappa$ and $\chi$ are used. The
functions $f$ and $g$ contain the information on major effects due
to the interaction with the gauge field and their explicit
expression can be found in Ref. \onlinecite{mdosy}. $Q_{0} =
\left( \kappa T/\chi \right)^{1/3}$ is a momentum cutoff and
$Q_0^{-1}$ can be identified as the length scale of gauge
fluctuations,  analogous to the anomalous skin depth. It is
important to note that, {\it w.r.t.} the PG estimate, the new
momentum scale $Q_0 \approx
\frac{{(1-\delta)}^{(2/3)}}{a}\left(\frac{6\pi T}{t}\right)^{1/3}$
is almost doping independent and is roughly bigger by a factor
$(\delta)^{-2/3}$ . If one integrates over  $|\vec x|$ in the
range $T^{-1}>> x^0
>> |\vec x|$ and makes use of the definition of $Q_0$ for the SM phase, the contribution of
the complex saddle point that dominated the integral in the PG
phase turns out to be small compared with the contribution coming
from fluctuations around the coordinate origin, in the region
$|\vec x|Q_0 \lesssim 1$,
 for $T$ or $\delta$ sufficiently large. Expanding $f$ and $g$ to leading
 order around the origin one derives:
\begin{equation} \label{om2}
\langle \vec \Omega(x) \cdot \vec \Omega(0) \rangle \simeq
\frac{1}{{x^0}^2}e^{-2im_sx^0+i\frac{T}{\chi m_s}\frac{|\vec
x|^2Q_0^2}{24} x^0} e^{-\frac{T}{2\chi}Q_0^2
g(0)\frac{{x^0}^2}{m_s^2}}.
\end{equation}
The integration over $|\vec x|$ is then simply Gaussian; defining
\begin{equation}\label{aa}
a=\frac{T}{\chi m_s}\frac{Q_0^2}{24}x^0
\end{equation}
and assuming $|a|\gg Q_0^2$, we are allowed to remove the cut-off
(the small real convergence factor needed can be
 supplied by the neglected higher order terms) obtaining (in the Fourier transformed representation
 for the space coordinates):
\begin{equation} \label{OO}
 \langle \vec \Omega \cdot \vec \Omega \rangle (x^0,\vec{q}=0) \simeq  \frac{1}{{x^0}^2}%
  e^{-2im_s x^0-\frac{T}{2\chi m_s^2}g(0) Q_0^2 {x^0}^2} \int_{|\vec x| {Q_0}\lesssim 1}
e^{i a {|\vec x|}^2}|\vec x|  d|\vec x| \simeq \frac{ i e^{-2im_s
x^0-\frac{T}{2\chi m_s^2}g(0) Q_0^2 {x^0}^2}}{a  {x^0}^2}.
\end{equation}
 The real part of the
exponent in (\ref{OO}) is monotonically decreasing in $x^0$,
therefore we estimate  the $x^0$ integral by principal part
evaluation. Since our approach is valid only at large scales, we
introduce an UV cutoff in the integral at $\lambda Q_0^{-1}$ and
evaluate the integration assuming $\lambda$ large. Then we make
the tentative conjecture that for small $\omega$ the physics is
dominated by large scales and the small-scale contribution can be
taken into account phenomenologically by a suitably chosen
rescaling of $\lambda$ (it cannot be sent to 0 as in the PG phase,
because here for
 $\lambda$=0 the exponent loses its convergence factor).
 The result of this approximation is:

\begin{equation} \label{ku-bo}
{\langle \vec \Omega \cdot \vec \Omega \rangle
(\omega,\vec{q}=0)}_{|\omega\rightarrow 0} \simeq \frac{\chi m_s
Q_0}{T } \frac{ e^{i (\omega-2m_s)\lambda Q_0^{-1}- \frac{T}{2
\chi m_s^2}g(0) \lambda^2 }}{\omega-2m_s+i\frac{T}{\chi m_s^2}g(0)
Q_0\lambda}.
\end{equation}

This result appears physically reasonable if the exponent in
(\ref{ku-bo}) is slowly varying and is of the order 1; this puts
limits, {\it a priori} on the validity of the above conjecture
given by:
\begin{eqnarray} \label{limit}
T &\gtrsim& (2 m_s \lambda)^3 \chi/\kappa \sim \lambda^3 t (\delta |\ln \delta|)^{3/2}, \nonumber \\
T &\lesssim&  2 \chi m_s^2 / (g(0)\lambda^2) \sim \lambda^{-2} t
\delta |\ln \delta|.
\end{eqnarray}
{\it A posteriori}, self-consistently $\lambda \sim O(1)$ (from
fitting $\lambda \approx 0.7$), and (\ref{limit}) selects a range
roughly between few tens and few hundreds Kelvin ($\approx
200-500$ for $\delta \approx 0.04-0.15$; for higher dopings the
upper limit of the temperature-dependent mass
treatment\cite{belk}
 $\approx 500 K$ applies). In the parameter range (\ref{limit}), equation (\ref{ku-bo})
 can be interpreted as follows:
the gauge fluctuations couple the spinon-antispinon pair into an
overdamped magnon resonance with mass gap
\begin{equation}
   m_\Omega= 2 m_s,
\end{equation}
inverse life-time
\begin{equation}
   \tau^{-1}_\Omega=\frac{T}{\chi m_s^2}g(0) Q_0 \lambda,
\end{equation}
and $T$-dependent wave-function renormalization factor $Z_\Omega
\sim \frac{\chi m_s Q_0}{T}$ . In this respect, the situation here
is qualitatively similar  to that appearing  in the PG phase, but
with different $T$ and $\delta$ dependence of the parameters.
Notice that typically in the SM phase $m_\Omega<<
\tau^{-1}_\Omega$ , whereas in the PG phase $(m_\Omega)_{PG}>>
(\tau^{-1}_\Omega)_{PG}$; that is to say, AF effect is more
pronounced in the PG phase than in the SM phase.

The Green's function $G$ for the 2D electron is given in real
space by the product of the holon and the spinon propagators,
averaged over gauge fluctuations. We are interested here in the
quasi-particle pole, in particular in the temperature and doping
dependence of the wave function renormalization constant $Z$ and
the damping rate $\Gamma$ defined via a representation of the
retarded electron correlation functions for  small $\omega$ and
momentum $\vec k_F$ on the Fermi surface as
\begin{equation} \label{G}
   G^R(\omega,\vec k_F) \sim \frac{Z}{\omega +i\Gamma}.
\end{equation}
To compute it we apply the dimensional reduction to the holon
propagator by means of the tomographic decomposition introduced by
Luther-Haldane.\cite{lu}  Following the procedure described in
Ref. \onlinecite{mdosy} we find
 within the range (\ref{limit}):
\begin{equation}\label{zetaSM}
   Z\approx \lambda_1 (\frac{Q_0}{k_F})^{\frac{1}{2}}
   (\frac{m_s\kappa}{J^2})^{\frac{1}{2}}.
\end{equation}
Comparing it with the result in Ref. \onlinecite{mdosy} one finds
that the anisotropic weight
 present in the PG phase has disappeared. As this corresponds to a restoration
 of the full Fermi surface, this suggests again that the above results apply only away
 from the boundary of the PG/SM crossover in a parameter region where
  ALL  effects due to PG have disappeared.  The electron damping rate $\Gamma$ at
the Fermi surface is:
\begin{equation}\label{dampSM}
 \Gamma=\frac{T}{2\chi m_s^2}Q_0 g(0)\lambda \propto T^{4/3}\frac{{(1-\delta)}^{5/3}}
 {\delta |\ln \delta|}.
\end{equation}
We see that $\Gamma$ is inversely proportional to the doping
concentration. This is qualitatively consistent with what was
found from the ARPES experiments, namely the quasi-particle peak
gets sharper with increasing doping.\cite{harris}

\section{Comparison with experiments}
 \subsection{In-plane Resistivity}
The in-plane resistivity is one of the most studied properties of
the High T$_c$ cuprates.  There is a narrow range of dopings
characterizing optimal/slightly overdoped samples where the
in-plane resistivity appear linear in $T$ from $T_c$  up to
several hundreds Kelvin. In underdoped samples, it deviates from
the linear dependence at low temperatures, first acquiring a
sublinear behavior and then crossing an inflection point (which we
take as the definition of the PG temperature $T^*$), with the
temperature dependence becoming almost quadratic. Finally, for
strongly underdoped samples it  reaches a minimum corresponding to
the MIC whose origin was discussed at length in Ref.
\onlinecite{mdosy}. In overdoped samples above a critical doping
concentration,  one observes again deviation from linearity at low
$T$, but there the behavior at low temperature is superlinear,
eventually at high doping reaching a Fermi liquid behavior $\sim
T^2$. With increasing doping, one finds the temperature of
deviation from linearity lowering in the underdoped region and
increasing in the overdoped region. This has been taken as one of
the experimental facts in favor of the existence of a QCP at
$\delta \sim 0.19$ in Ref. \onlinecite{loram}.

To calculate the in-plane resistivity we apply the Ioffe-Larkin
rule,\cite{Iof}
\begin{equation}
   \rho= \rho_s + \rho_h.
\label{il}\end{equation} Using the Kubo formula to calculate the
spinon contribution to conductivity, within the range
(\ref{limit})
 one finds \begin{equation}\label{asympt}
\rho_s\simeq 2 \lambda Q_0^{-1} (\Gamma +  m_s^2/ \Gamma)
= \frac{T}{\chi m_s^2}g(0)\lambda^2 + \frac{4 m_s^4 \chi}{T g(0) Q_0^2}. %
\end{equation}
In the high temperature limit $Q_0 \gg m_s$, the damping rate in
(\ref{asympt}) dominates over the spin gap $2m_s$ and the spinon
contribution to resistivity is linear in $T$, with a slope $\alpha
= g(0)\lambda^2/(\chi m_s^2) \simeq (1-\delta)/(\delta |\ln
\delta|)$.
 A  linear in $T$ behavior is also obtained in the gauge field theory of Nagaosa and Lee\cite{Lee}
for the uniform RVB state with a very similar slope $\alpha_{RVB}
\simeq 1/ \delta$. Lowering the temperature, the second term in
(\ref{asympt}) gives rise first to a superlinear behavior and
then,  at the margin of validity (\ref{limit}) of our approach, an
unphysical upturn. The deviation from linearity is due to the spin
gap effects and is cutoff in the underdoped samples by the
crossover to the PG phase. We expect that physically in the
overdoped samples it is cutoff  by a crossover to a Fermi liquid
``phase''.

We now turn to holon contributions. It is known since Nagaosa and
Lee,\cite{Lee} that for a 2D Fermi gas scattering against a U(1)
gauge field with Reizer-like singularity, the  scattering time at
the Fermi surface is proportional to $T^{-4/3}$. This power law
follows simply from the scaling analysis and does not depend on
the detail of the dispersion relations. The contribution $\rho_h$
from our gas of spinless holons is of the form
\begin{equation}\label{hhh}
\rho_h \sim \left(\frac{1}{\tau_{imp}}+\epsilon_F(\frac{T}{\epsilon_F})^{4/3}\right),
\end{equation}
where we also added  the contribution of the  impurity scattering
via the Matthiessen rule. Compared with (\ref{asympt}), it gives a
subleading contribution.  This could explain, as for the PG phase,
the insensitivity of the resistivity to the presence of
non-magnetic impurities which affect only $\rho_h$.
 The above results reproduce
qualitatively the $T-$linearity of the in-plane resistivity in the
SM phase , {\it including the decrease of the slope upon doping
increase} and the superlinear behavior at low $T$ for overdoped
samples. One can use the deviation from linearity appearing in the
 overdoped samples to fix the phenomenological parameter $\lambda$ at some doping by computing
 the scale independent quantity $(\rho(T)-\rho_h(0))/\alpha$ and comparing it with experimental data.
 The result is self-consistently determined as $\lambda$ of the order 1, and more precisely, using LSCO data
 ( for $\delta =0.30$\cite{sundqvist}) one finds
 $\lambda \approx 0.7$  as quoted above. The temperature dependence of the spinon mass yields
  a bending at high temperature, stronger for lower dopings, as visible in the resistivity data
   at constant volume for LSCO;\cite{sundqvist} this effect is masked in the resistivity data at
    constant pressure by thermal expansion.

\begin{figure}
\begin{center}
\includegraphics[height=8cm,angle=0]{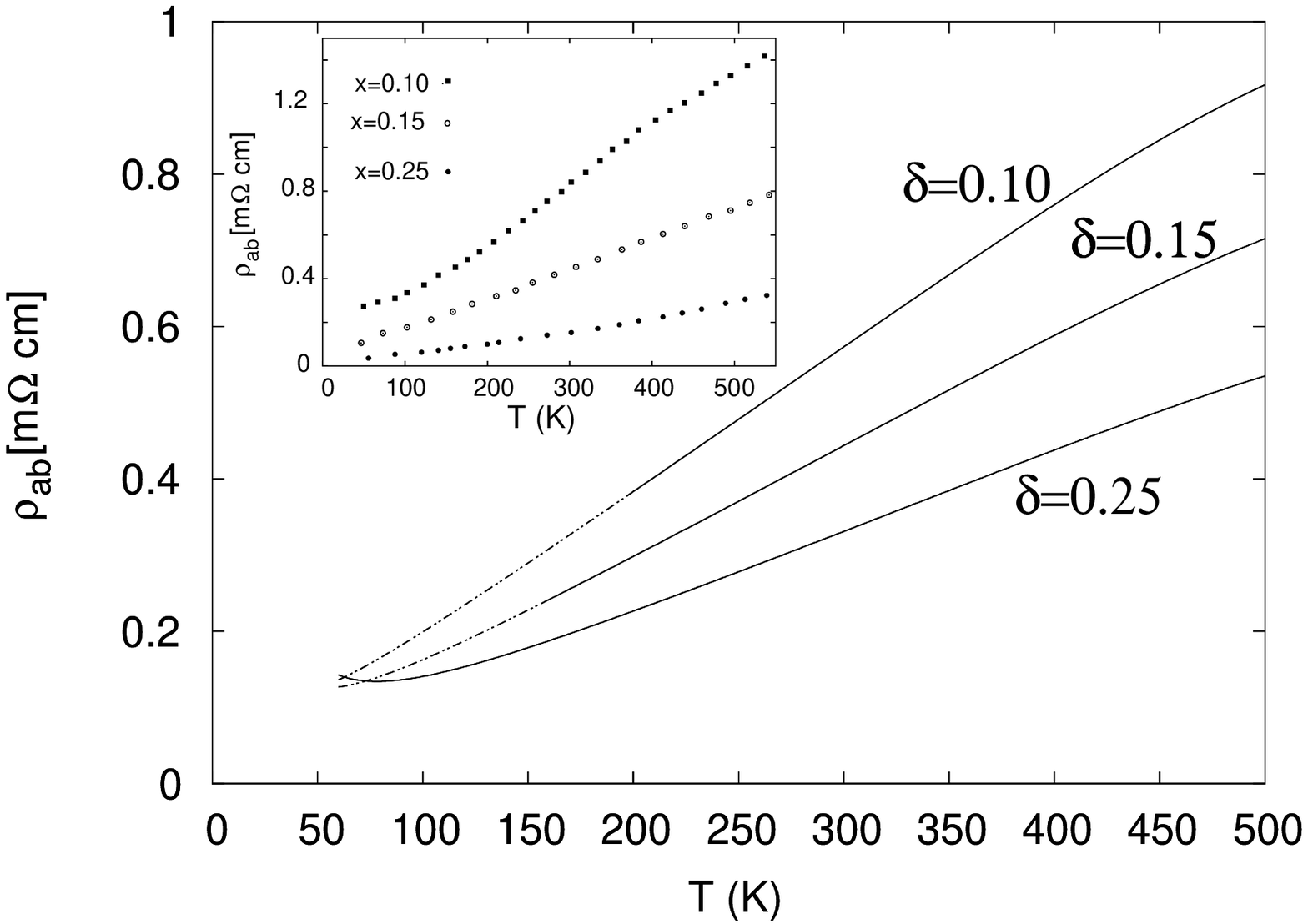}
\caption{Calculated in-plane resistivity as a function of
temperature for different hole concentrations.  Below the
pseudo-gap temperature $T^*$, the curve is shown in dashed line.
The resistivity scale is fixed by comparison with experimental
data for $\delta =0.15$ at 400 K.  {\sl Inset:} In-plane
resisitivity versus $T$ measured in LSCO crystals with different
Sr content $x$, taken from the work of Takenaka {\it et al.}, Ref.
\onlinecite{sundqvist}}.\label{rab}
\end{center}
\end{figure}

In Fig.~\ref{rab} we plot the in-plane resistivity given by
Eq.(\ref{asympt}) versus temperature for different dopings. Except
for an overall scale, once $\lambda$ is fixed, there are no
additional free parameters for other doping concentrations in
Eq.(\ref{asympt}) and the agreement with experiments is reasonably
good.

We should mention that the linear temperature dependence of the
in-plane resistivity  is also reproduced in a number of
theoretical studies, including  the ``marginal Fermi liquid''
theory,\cite{varma89} the ``hot spots'',\cite{stojkovic,abanov} as
well as ``cold spots''\cite{io-mi} approaches, although the
mechanism leading to the linear dependence varies from case to
case, and is also somewhat different from our  consideration. We
should also mention that since we ignore all details of the Fermi
surface, some properties, like the Hall effect, strongly depending
on these details, cannot be treated properly at this stage in our
approach. This requires further studies.

\subsection{Spin-Lattice Relaxation Rate}
From the experimental point of view, one of the hallmarks of the
SM phase is the unusually  simple law for the spin-lattice
relaxation rate at the Cu-sites:
\begin{equation}
(\frac{1}{T_1T})^{63}\sim \lim_{\omega \rightarrow 0} \sum_{\vec q}
\frac{{\rm Im}\chi_s(\omega,\vec q)}{\omega} F(\vec q)\sim\frac{1}{T}.
\end{equation}
where $F$ is the hyperfine formfactor peaked around the AF wave
vector ${\cal Q}_{AF}$. While in optimally doped samples the above
proportionality relation is valid over a wide range of
temperatures above $T_c$, in overdoped samples
$(\frac{1}{T_1T})^{63} $ saturates to a constant (i.e.
$T-$independent) value at low temperatures, suggesting a possible
crossover to a Fermi liquid phase. Using the expression
(\ref{om2}) for the magnon correlator one obtains a factorized
form for the spin susceptibility: $\chi_s(\omega, \vec q) \sim
\chi(\omega, \vec q=0) \Xi(\vec q)$ with $\int d^2 \vec q \Xi(\vec
q)=1$, a feature present also in the PG phase and that  has been
claimed to be in agreement with experimental
data.\cite{littlewood}

Then in our computation we are left with the following integral:
\begin{equation}
\label{tt163}
(\frac{1}{T_1T})^{63} \sim (1-\delta)^2 \int dx^0 ix^0 \frac{1}{{x^0}^2}
e^{-2im_s x^0-\frac{T}{2\chi m_s^2}g(0) Q_0^2 {x^0}^2}.
\end{equation}
Up to the factor $(1-\delta)^2$, the above integral is equal to
the spinon contribution to the conductivity.

\begin{figure}
\begin{center}
\includegraphics[height=8cm,angle=0]{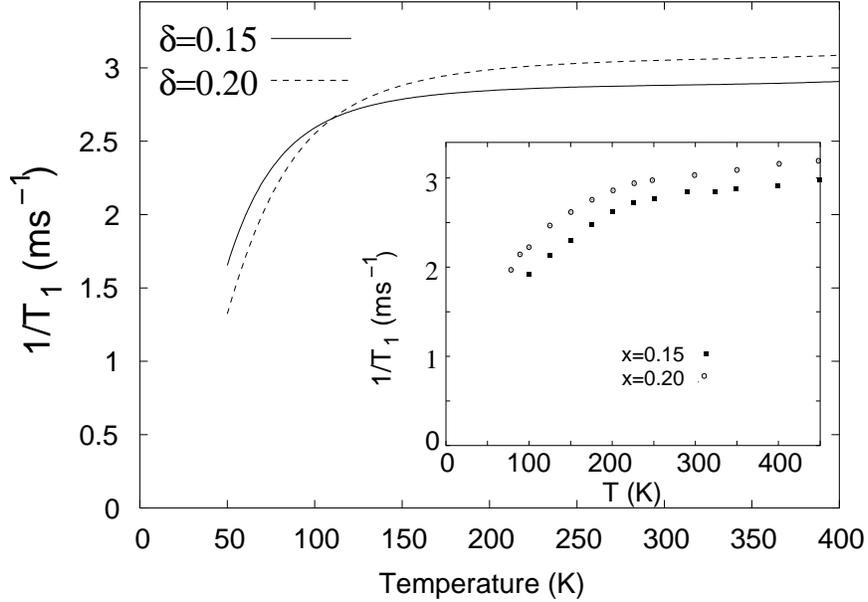}
\caption{Calculated temperature dependence of the spin-lattice
relaxation rate $1/T_1$ for different doping concentrations. {\sl
Inset:} spin-lattice relaxation rate measured in LSCO samples with
different Sr content $x$, taken from the work of Fujiyama {\it et
al.}, Ref.  \onlinecite{berthier}.} \label{spi}
\end{center}
\end{figure}

Therefore, in the high temperature limit we recover the linear in
$T$ behavior for
 $(T_1T)^{63} \approx   T /((1-\delta)^2 \chi m_s^2)$, while at high dopings and low
 temperatures
 the superlinear deviation,  also found experimentally in overdoped
samples of LSCO.\cite{berthier}
 Furthermore, the factor $(1-\delta)^{-2}$ weakens the doping dependence of the slope as compared
 with the resistivity curves, in agreement with the experimental data. In Fig.~\ref{spi}
  we plot $1/T_1^{63}$ extracted from Eq.(\ref{tt163})  versus temperature
for different dopings.

 \subsection{Out-of-plane Resistivity}
Conductivity along the c-axis in cuprates appears as mainly due to
the interlayer tunneling process. Since there is no measurable
Fermi velocity along this axis because of very small (effective
\cite{note}) hopping integral $t_c$, we can use Kumar-Jayannavar
(K-J) approach to calculate $\rho_c (T)$. According to these
authors,\cite{Kumar} the out-of-plane motion in cuprates is
incoherent and governed by the strong in-plane scattering via the
quantum blocking effect. At high temperatures $\Gamma \gg t_c$ the
interlayer tunnelling rate is reduced by the in-plane scattering.
Under these assumptions, $\rho_c (T)$ is controlled by the second
term in the K-J formula:
\begin{equation} \label{kuma}
   \rho_{c} \sim \frac{1}{\nu}\left(\frac{1}
   {\Gamma}+\frac{\Gamma}{t_c^2 Z^2}\right).
\end{equation}
In the SM phase,
substituting $\Gamma$ and $Z$ with the corresponding estimates (\ref{dampSM}) and
(\ref{zetaSM}) in (\ref{kuma}), we recover the $T-$linearity in the
``incoherent regime'' $\Gamma \gg t_c Z$:
\begin{equation}\label{rc}
\rho_c \simeq \frac{J^2}{t_c^2}\frac{k_F}{m_s
\nu(\epsilon_F)\kappa}\frac{T}{2\chi m_s^2}\simeq
\frac{T}{(\delta|\ln\delta|)^{3/2}} \end{equation}
 decreasing
faster than $\rho_{ab}(T)/T$ upon doping increase.

  The out-of-plane resistivity was also
calculated in a number of theoretical studies, including the gauge
field approach\cite{nagaosa} and ``cold spots'' model.\cite{io-mi}
As far as we understand, the correct temperature dependence in the
SM phase could not be reproduced there.

\begin{figure}
\begin{center}
\includegraphics[height=8cm,angle=0]{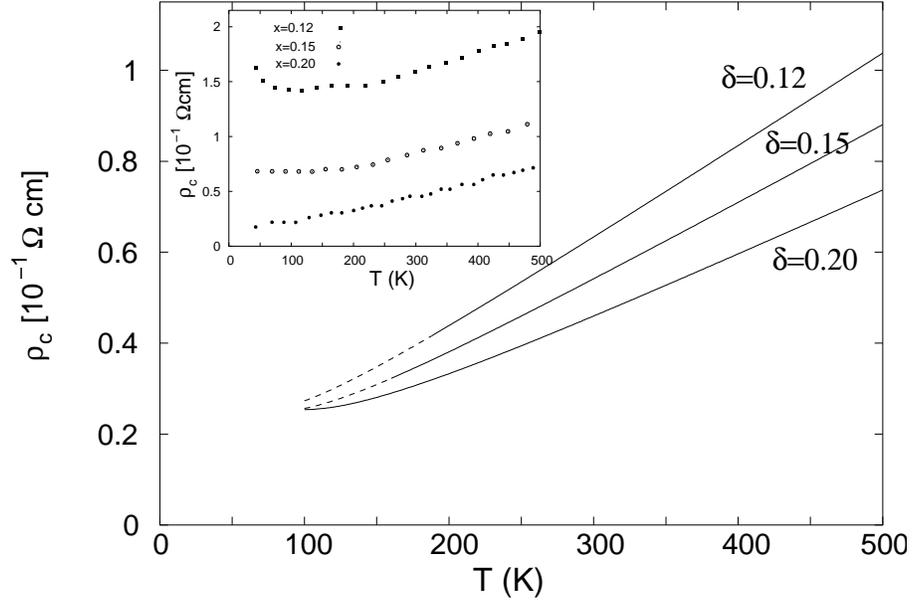}
\caption{Calculated out-of-plane resistivity as a function of
temperature for different hole concentrations. Below the
pseudo-gap temperature $T^*$, the curve is shown in dashed line.
{\sl Inset:} In-plane resistivity versus $T$ measured in LSCO
crystals with different Sr content $x$, taken from the work of
Nakamura {\it et al.}, Ref.  \onlinecite{nakamura}}.
\label{rc-fig}
\end{center}
\end{figure}

Equation (\ref{kuma}) would predict a minimum in $\rho_c$ at low
$T$, unless it is cutoff by a crossover to a new ``phase''. Hence,
in the spin-charge gauge approach $\rho_c$ might exhibit a MIC for
three different reasons, each one with its own $T$ and $\delta$
dependence: 1) a K-J minimum in the PG phase; 2) a minimum due to
a crossover from an insulating regime in the PG phase to a
metallic regime in the SM phase; 3) a K-J minimum in the SM phase.
Most of the MIC in $\rho_c$ exhibited in the experimental data
appear to correspond to the second case; thus the minimum roughly
corresponds to the deviation from linearity. The first case is
usually cutoff by the crossover to the SM phase. Perhaps an
example of the third case is the minimum found in BSCO at $\delta
\approx 0.225$ \cite{krusin} suppressing superconductivity with a
magnetic field, as the minimum there is lower than the deviation
from linearity. However,  the parameter region where this minimum
has been found is close to the boundary of the SM phase, so one
could expect corrections to our formulas that has to be
investigated. In Fig.~\ref{rc-fig} we plot $\rho_c$ extracted from
Eq.(\ref{kuma}) versus temperature for different doping
concentrations and a fixed value of the relative coefficient of
the metallic versus insulating term, $r \sim J^2/(t_c^2
\lambda_1^2)$ (with $\lambda_1$ as  given in Eq.(\ref{zetaSM}))
and compare the result with data in LSCO.\cite{nakamura}

\begin{figure}
\begin{center}
\includegraphics[height=8cm,angle=0]{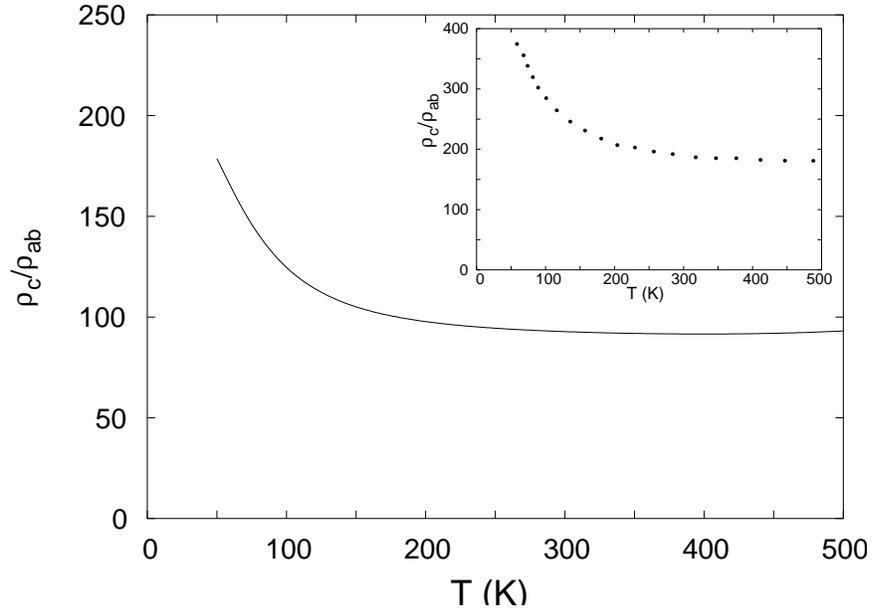}
\caption{Calculated resistivity anisotropy  ratio as a function of
temperature for fixed hole density $\delta=0.2$. {\sl Inset:}
resistivity anisotropy ratio measured for a LSCO sample with Sr
content $x=0.20$, taken from the work of Nakamura {\it et al.},
Ref.  \onlinecite{nakamura}.} \label{ratio}
\end{center}
\end{figure}

Unfortunately, we don't have a reliable method to estimate the
(extrapolated) $T=0$ intercept of $\rho_c(T)$ which is large when
compared with the corresponding intercept for  $ab-$plane
resistivity, and it is also difficult to fix precisely using
experimental data the relative coefficient $r$, so we cannot
extract safely the anisotropy ratio $\rho_c (T)/\rho_{ab} (T)$.
However, let us notice that if the minimum in $\rho_c$ is at
higher temperature than the (unphysical) minimum in $\rho_{ab}$,
then one can derive the fast decrease of the anisotropy ratio at
low temperature found experimentally,\cite{nakamura} see
Fig.\ref{ratio}.

Finally let us propose a simple qualitative explanation of the
linear $T$-behavior of both in-plane and out-of-plane resistivity
in the ``interior"'' of the SM phase in terms of ``effectiveness''
in the momentum space, somewhat analogous to the ``effectiveness''
appearing, in the coordinate space, in the treatment of anomalous
skin effect (see e.g. Ref. \onlinecite{callaway}). In our approach
the electron resonance life-time was found to be $\tau_e \sim T^{-
\frac{4}{3}} $, so that from the Boltzmann transport formula
naively one expects to find a conductivity $\sigma_0 \sim T^{-
\frac{4}{3}}$. However the gauge field is able to combine spinon
and holon into a resonance only in a range of momenta of the order
of the anomalous skin momentum $Q_0 \sim T^{\frac{1}{3}}$, so only
a fraction  of electrons $Q_0/p_F$ should contribute or be
``effective'' for conductivity, so that $\sigma \sim \sigma_0 Q_0
\sim T^{- \frac{4}{3}} T^{\frac{1}{3}}= T^{-1}$.

\subsection{AC conductivity}
\begin{figure}
\begin{center}
\includegraphics[height=8cm,angle=0]{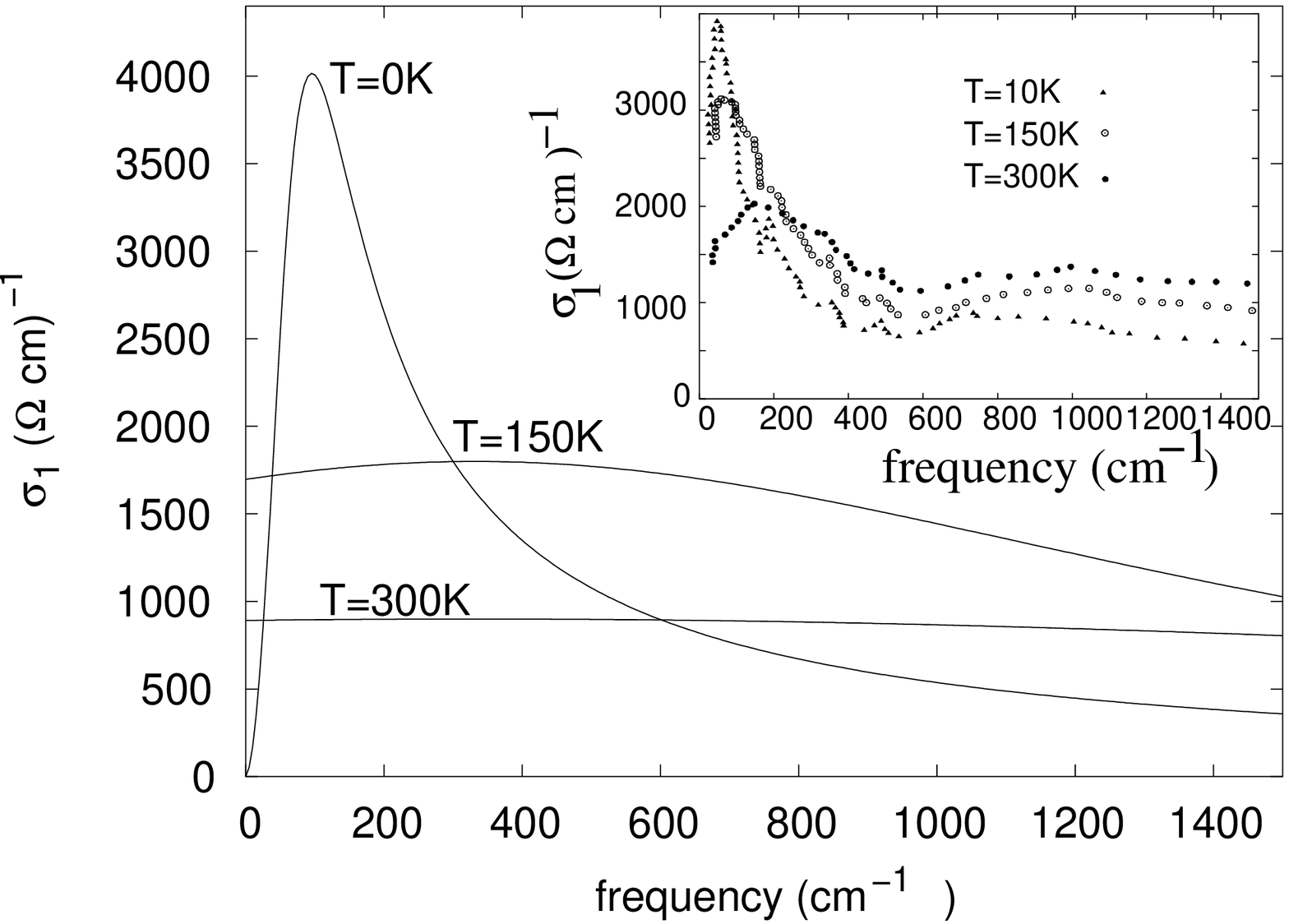}
\caption{Calculated AC conductivity as a function of frequency for
fixed hole density $\delta=0.184$ and different temperatures. {\sl
Inset:} AC conductivity versus frequency measured for a LSCO
sample with $x=0.184$ at different temperatures, taken from the
work of  Startseva {\it et al.}, Ref. \onlinecite{startseva}.}
\label{ac}
\end{center}
\end{figure}

A key feature of AC conductivity in the SM phase is a high
frequency tail $\sim \omega^{-1}$, found already in earlier
experiments. However recently it has been observed (in LSCO
\cite{startseva}
 and BSCO \cite{lupi}) that it is related to an asymmetric  peak structure in overdoped samples
  centered at a finite frequency, increasing with doping, shifting to higher frequency and
  symmetrizing when the temperature increases beyond the peak frequency at low $T$.
All these features are qualitatively reproduced in our spin-charge gauge
approach. As explained in Ref. \onlinecite{anis}, to compute the
AC conductivity one first evaluates the thermal
 gauge propagator with UV cutoff in the $\omega$-integration given by  $\Lambda=$max($T,\Omega)$,
 where $\Omega$ is the external frequency and as usual the thermal function coth$\omega/T$ is
  approximated by $T/\omega$ and sign$\omega$ in the two limits $\Omega << T$,$T <<
  \Omega$,
  respectively. It turns out that up to the logarithmic accuracy one can pass from the first
  to the second limit by replacing  $T$ with $\Omega$ in $Q_0$ and $\Gamma$ ( we denote the
  obtained quantities by $Q_\Omega, \Gamma_\Omega$) and rescaling the $f$ and $g$ functions
  resulting from the gauge field integration, by a positive multiplicative factor
  $\tilde\lambda \lesssim 1/2$. It is straightforward  then to derive the leading spinon
  contribution to the AC conductivity at finite T in the two limits.
For   $\Omega << T$ we have
\begin{equation}
\sigma(\Omega,T) \sim \frac{Q_0}{i (\Omega -2 m_s) +\Gamma} \sim \frac{1}{i (\Omega -2 m_s)
 T^{-1/3} + T},
\end{equation}
while for $T << \Omega$
\begin{equation}
\sigma(\Omega,T) \sim \frac{Q_\Omega}{i (\Omega -2 m_s)+ \tilde\lambda \Gamma_\Omega}
 \sim \frac{1}{i (\Omega -2 m_s) \Omega^{-1/3} + \Omega}.
\end{equation}
 From the above formulas the features discussed at the beginning of
this subsection
 are easily derived: the tail $\sim \Omega^{-1}$ is evident and the effect of replacing  $Q_0$ by $Q_\Omega$
 is to asymmetrize the peak at $2 m_s$ appearing in Re$\sigma(\Omega,T)$ for $\Omega << T$
 and to shift it towards lower frequency. The optical conductivity
 was discussed in a number of theoretical studies, including the
 spin-fermion approach,\cite{abanov} ``nearly AF Fermi liquid''
 theory\cite{stojkovic} and ``cold spots'' model.\cite{io-mi}
 To the best of our knowledge, the above considered features were
 not addressed earlier.
In Fig.~\ref{ac} we plot the AC conductivity in the two regimes
considered above at fixed doping. Inclusion of the contribution of
holons via Ioffe-Larkin rule does not change the qualitative
features, but enhances the finite temperature curves, improving
the agreement with experiments.

Two remarks are in order. First, our formulas do not apply for
$\Lambda$ close to zero, where we expect a transition to a
different ``phase'' as discussed above. Second, one can prove that
the peak seen in the SM phase is replaced in the PG phase by a
broad maximum (moving to lower frequency as doping increases), due
to the ``relativistic'' structure of the spinon peculiar to that
phase, as discussed in Ref.  \onlinecite{anis}.

\section{concluding remarks}

To conclude we have shown that the spin-charge gauge approach
originally proposed to describe the PG phase of the cuprate
superconductors, in particular, the MIC in the underdoped cuprates
can be generalized to the SM phase as well. The common features
are the interplay of the AF short range order and the gauge field
fluctuations, which leads to magnon and electron resonances, while
the difference is the presence of a statistical $\pi$-flux lattice
for holons in the PG phase responsible for the small
``Fermi-arcs'' and its absence in the SM phase.  As the
doping/temperature increases the MIC is taking place, and then the
flux lattice ``melts''.  It is interesting to observe that the
extrapolations from the PG and SM regions more or less match each
other in the crossover area. In some sense the PG treatment is
doing a ``better job'', producing the MIC and sublinear
temperature dependence for the in-plane resistivity, up to the
inflection point $T^*$ as the margin of the PG phase. It is true
that the ``high temperature asymptotics'' $\sim T^{1/4}$ derived
in the PG phase does not correspond to experiment, but that has
been corrected by the SM consideration and the disappearance of
the ``Fermi-arcs'' feature allows to recover the celebrated linear
in $T$ dependence. This dependence in our approach can be
interpreted as due to the combined effect of an electron life time
$\sim T^{-\frac{4}{3}}$ due to gauge interactions triggered by
Reizer singularity, and the ``effective'' slab in momentum space
where the electron resonance forms $\sim Q_0 \sim
T^{\frac{1}{3}}$, with $Q_0$ as the anomalous skin momenta again
associated with the Reizer singularity.

In our view there seems to be a crossover line between the
`insulating' and `metallic' regions at any finite temperature.
Probably, this crossover becomes a quantum phase transition at
zero temperature between the AF short-range ordered state and
Fermi liquid state, although the transition point seems not being
able to control the behavior of the system over a large region,
because of the appearance of new scales (AF correlation length)
nearby.  This scenario seems  consistent with the picture
extracted from the $\mu$SR measurements in the SC
state.\cite{pana} The phase diagram that our approach suggests is
in a sense intermediate between the RVB and the QCP picture,
sharing with
 the first scenario the nature of crossover for the border between the
 PG and SM phases at finite $T$, while with the second scenario the existence
 of a true phase transition at $T=0$.

\appendix*\section{}

 In this Appendix, using Ref. \onlinecite{Mar} as a guideline,
we show how to satisfy on average the {\it Assumption} 0. We split
the integration over $V$ into an integration over a field
$V^{(c)}$, satisfying the Coulomb condition:

\begin{equation}
\partial^\mu V_\mu^{(c)} =0,
\label{su2cou}
\end{equation}
and its gauge transformations expressed in terms of an $SU(2)$--valued
scalar field $g$ , i.e.,
$V_a =g^\dagger V_a^{(c)} g + g^\dagger\partial_a g, \; a = 0,1,2.$

Integrating over the time component of the C-S gauge fields one finds:

\begin{equation}
B_\mu = \bar B_\mu + \delta B_\mu,\;\;\; \delta B_\mu (x) = \frac{1}{2}
 \sum_j H^*_j H_j \partial_\mu {\rm arg} \ (x-j),
\label{u1fix}
\end{equation}
where $\bar B_\mu$ gives rise to a $\pi$-flux phase, i.e.,
$e^{i\int_{\partial p} \bar B} = -1$ for every plaquette $p$ and

\begin{equation}
V_\mu^{(c)} = \sum_j (1- H^*_j H_j) (\sigma_x^{|j|} g^\dagger_j
\frac{\sigma_a}{2} g_j \sigma^{|j|}_x)_{11} \partial_\mu {\rm arg} \
(x-j) \sigma_a,
\label{su2fix}
\end{equation}
where $\sigma_a, a= x, y, z$ are the Pauli matrices. It is not
difficult to see that the statistical field $V^{(c)}$ defined in
(\ref{su2fix}) does not carry flux $\pi$ per plaquette, so that
strictly speaking the constraint ${\rm arg}  ( U_{\partial p}) =
0$ cannot be satisfied. However, it is sufficient for our purposes
that whenever a hole make a closed loop on the lattice, it
acquires a trivial phase factor $(2 \pi n)$.

At the lattice level,
the particle can either stay at rest at a given site or jump to a nearest neighbor.

Without loss of generality, we consider a holon going around a given
plaquette $p$ counterclockwise.

Now assume that the particle hops from site $i$ to site $j$. Let
${\underline{\omega}}$ denote the set of trajectories of holons in
a first-quantization path representation. In such formalism
 the trajectories are left continuous in time, i e. one should think of the holon in the link
 located at the end of the jump. It is shown in Ref \onlinecite{Mar} that for the
 optimal configuration, independently of $\delta$,

 \begin{equation}\label{cox }
\Bigl(\sigma_x^{|i|} g^\dagger_i e^{ i\int_{<ij>} V^{(c)}} g_j
\sigma_x^{|j|}\Bigr)_{11} \simeq \Bigl(\sigma_x^{|i|} \tilde g^\dagger_i
\tilde g_j \sigma_x^{|j|}\Bigr)_{11}.
\end{equation}

where the SU(2) variables $\tilde g_i, \tilde g_j$ have the structure:
\begin{equation}
\tilde g_i=e^{i\theta_i \sigma_z},\;\;\;\;i \notin
{\underline{\omega}}
\end{equation}
\begin{equation}
\tilde g_j=i(\vec \sigma \cdot \vec n_j),\;\;\;n_j=(\cos \phi_j,\sin \phi_j,0),\;\;\;j \in {\underline{\omega}}
\end{equation}
with $\theta_j$ and $\phi_j$ so far arbitrary angles.

These degrees of freedom will
be fixed in order to cancel the $\pi-$magnetic flux per plaquette generated
by $\bar B$. We need to impose:
\begin{equation}\label{conn}
\prod_{<ij> \in \partial p} \Bigl(\sigma_x^{|i|} \tilde g^\dagger_i
\tilde g_j \sigma_x^{|j|}\Bigr)_{11}=e^{i \pi}.
\end{equation}

Writing out explicitly the left-hand side of (~\ref{conn}), we
find
\begin{equation}
e^{i(\theta_1+\phi_2)} e^{-i(\theta_2+\phi_4)} e^{i(\theta_4+\phi_3)}
e^{-i(\theta_4+\phi_1)}
\end{equation}
which  is satisfied by the choice $\phi_j=\frac{\pi}{4}(-1)^{|j|}$
irrespective of the $\theta_j$ phases, provided the latter
contributes  with a trivial phase factor.

Using this gauge freedom we can be even more demanding, by requiring that the field $\bar B$ is
exactly cancelled link by link. For instance, imposing
$$
\theta_j=-\frac{\pi}{2}(-1)^{|j|},
$$
we cancel the distribution of phase factors for $\bar B$ chosen in
Ref \onlinecite{Mar}. This proves that choosing the SU(2) gauge
field as described above, the {\it Assumption} 0 is satisfied.

{\bf Acknowlegments.} One of us (P.M.) wishes to thank C. Pepin,
G. Qin and T. Zillio for useful discussions and the Chinese
Academy of Sciences for kind hospitality. Interesting discussions
with T. Xiang are gratefully acknowledged.

\begin{thebibliography}{9}
\bibitem{norman}See, {\it e.g.}, the recent review M.R. Norman and C. P\'{e}pin,  Rep. Progr. Phys. {\bf 66},
1547(2003).
\bibitem{bas} G. Baskaran, Z. Zou, and P.W. Anderson, Solid State Commun. {\bf 63}, 973 (1987);
G. Kotliar and J. Liu, Phys. Rev. B {\bf 38}, 5142 (1988); H.
Fukuyama, Progr. Theor. Phys. Suppl. {\bf 108}, 287 (1992); N.
Nagaosa and P.A. Lee, Phys. Rev. B {\bf 45}, 966 (1992).
\bibitem{sachdev} See, {\it e.g.}, S. Sachdev, ``Quantum Phase Transitions'', Cambridge University Press, 1999.
\bibitem{sachdev-rmp} S. Sachdev, Rev. Mod. Phys. {\bf 75}, 913 (2003).
\bibitem{kivelson} S.A. Kivelson {\it et al.}, Rev. Mod. Phys. {\bf 75}, 1201 (2003).
\bibitem{varma} C. M. Varma \prb {\bf 55}, 14554 (1997); S. Chakravarty,
R.B. Laughlin, D. Morr, and C. Nayak \prb {\bf 63}, 094503 (2001).
\bibitem{castellani} C. Castellani {\it et al.}, Z. Phys. B {\bf 103}, 137 (1997);
J. Phys. Chem. Solids {\bf 59}, 1694 (1998).
\bibitem{pines} D. Pines, Z. Phys. B {\bf 103}, 120 (1997) and references therein.
\bibitem{abanov} Ar. Abanov, A.V. Chubukov, and J. Schmalian, Adv.  Phys. {\bf 52}, 119 (2003)
(See cond-mat/0107421).
\bibitem{loram} J.L. Tallon {\it et al.}, Phys. Stat. Sol. (b) {\bf
215}, 531 (1999); J.L. Tallon {\it et al.}, cond-mat/0211048.
\bibitem{Mar} P.A. Marchetti, Z.B. Su, L. Yu,
Phys. Rev. B {\bf 58},  5808 (1998).
\bibitem{Dai} P.A. Marchetti, J.H. Dai, Z.B. Su and L.Yu, J. Phys. Condens. Matt. {\bf 12}, L329
(2000).
\bibitem{magneto} P.A. Marchetti, Z.B. Su and L. Yu, Phys. Rev. Lett. {\bf 86}, 3831 (2001).
\bibitem{mdosy} P.A.Marchetti, L. De Leo, G.Orso, Z.B. Su
and Yu Lu, Phys. Rev. B {\bf 69} 024527 (2004).
\bibitem{anis} P.A.Marchetti, G.Orso, Z.B. Su and Yu Lu,  Phys. Rev. B {\bf 69}, 214514 (2004).
\bibitem{largeFS} A. Damascelli, Z. Hussain, and Z.X. Shen, Rev. Mod. Phys. {\bf 75}, 473
(2003).
\bibitem{fukuyama} See, {\it e.g.}, H. Fukuyama, Progr. Theor. Phys. Suppl. {\bf 108}, 287 (1992).
\bibitem{Lee} P.A. Lee and N. Nagaosa, Phys. Rev. Lett. {\bf 65}, 2450  (1990); Phys. Rev. B {\bf 46},
5621(1992).
\bibitem{iowi}  L.B. Ioffe and P.B. Wiegmann, Phys. Rev. Lett. {\bf
65}, 653 (1990); L.B. Ioffe and G. Kotliar, \prb {\bf 42}, 10348
(1990).
\bibitem{ichinose} I. Ichinose {\it et al.}, Phys. Rev. B {\bf 64},
104516 (2001).
\bibitem{stojkovic} B.P. Stojkovic and D. Pines, \prb {\bf 55}, 8576, {\bf 56}, 11931 (1997).
\bibitem{io-mi} L.B. Ioffe and A.J. Millis, \prb {\bf 58}, 11631
(1998); see, also, A.T. Zheleznyak {\it et al.}, \prb {\bf 57},
3089 (1998).
\bibitem{Fro} J. Fr\"{o}hlich and P.A. Marchetti, Phys. Rev. B
{\bf 46},  6535 (1992).
\bibitem{Birgenau} R. T. Birgeneau {\it et al.}, Phys. Rev. B {\bf 38}, 6614 (1988).
\bibitem{Reizer} M. Reizer, Phys. Rev. B {\bf 39}, 1602
(1989); {\bf 40},  11571 (1989).
\bibitem{lieb} E.H. Lieb, Phys. Rev. Lett.{\bf 73}, 2158 (1994).
\bibitem{rammal}J. Bellisard and R. Rammal, Europhys. Lett.  {\bf 13}, 205 (1990).
\bibitem{qin} G. Qin, unpublished.
\bibitem{tree} This identification is true only at the tree level. Since the gauge field propagator is changed,
the renormalized spinon propagator in the SM phase will be quite
different from that for the PG phase. One should also take note
that the mass term here $m_S$ is due to vortices, and it is not
the mass term in the quantum disordered phase of the standard
CP$^1$ model.\cite{sachdev}
\bibitem{belk} B. Keimer {\it et al.}, Phys. Rev. B {\bf 46}, 14034 (1992).
\bibitem{orso} G. Orso, Ph. D. Thesis, SISSA, 2003.
\bibitem{lu} A. Luther, Phys. Rev. B {\bf 19}, 320  (1979); F.D.M.
Haldane, Varenna Lectures 1992.
\bibitem{harris} J.M. Harris {\it et al.},\prl {bf 79}, 143 (1997).
\bibitem{Iof} L. Ioffe and A. Larkin, Phys. Rev. B {\bf 39},  8988 (1989).
\bibitem{sundqvist} B. Sundqvist and E.M.C. Nilsson, Phys. Rev. B {\bf 51}, 6111 (1995);
 K. Takenaka {\it et al.}, Phys. Rev. B {\bf 68}, 134501 (2003).
\bibitem{varma89} C.M. Varma {\it et al.}, \prl {\bf 63}, 1996 (1989).
\bibitem{littlewood} P.B. Littlewood, in ``Strongly interacting fermions and High
$T_c$ superconductivity'', Proceedings, Les Houches 1991.
\bibitem{berthier} C. Berthier {\it et al.},  J. Physique I {\bf 6}, 2205 (1997); S. Fujiyama {\it
et al.},  J. Phys. Soc. Jpn. {\bf 66}, 2864 (1997).
\bibitem{note} For simplicity we neglect the $k$-dependence of the hopping integral,
replacing it by an averaged effective constant hopping parameter
$t_c$.
\bibitem{Kumar} N. Kumar and A.M. Jayannavar, Phys. Rev.  B {\bf 45}, 5001 (1992);
 N. Kumar {\it et al.}, Mod. Phys. Lett. B {\bf 11}, 347  (1997); Phys. Rev. B {\bf 57}, 13399
(1998).
\bibitem{nagaosa} N. Nagaosa, \prb {\bf 52}, 10561 (1995).
\bibitem{krusin} L. Krusin Elbaum {\it et al.},  Int. J. Mod. Phys. B
{\bf 17}, 3598 (2003).
\bibitem{nakamura} Y. Nakamura and S. Uchida, Phys. Rev. B {\bf 47}, 8369 (1993).
\bibitem{callaway} J Callaway, ``Quantum Theory of the Solid State'', Academic Press, 1991
\bibitem{startseva} T. Startseva {\it et al.}, Physica C {\bf 321}, 135 (1999).
\bibitem{lupi} S. Lupi {\it et al.},  Phys. Rev. B {\bf 62}, 12418 (2000).
\bibitem{pana}  C. Panagopoulos {\it et al.}, Solid State Commun. {\bf 126}, 47 (2003)
\end{thebibliography}
\end{document}